\documentclass[12pt]{article}
\usepackage[a4paper]{geometry}
\usepackage{enumerate}
\usepackage{mathtools}
\usepackage{amsmath}
\usepackage{graphicx}
\usepackage{epsfig}
\usepackage{sidecap}
\usepackage{enumerate}
\usepackage{hyperref,stackrel}
\usepackage[normalem]{ulem}
\usepackage[bottom]{footmisc}
\usepackage{color}
\usepackage{enumerate}
\usepackage{bm}
\usepackage[utf8]{inputenc}
\usepackage{amsmath}
\usepackage{amsfonts}
\usepackage{amssymb}
\usepackage{graphicx}
\usepackage{enumitem}
\usepackage{authblk}
\usepackage{hyperref}
\usepackage [english]{babel}
\usepackage{color}
\usepackage[autostyle]{csquotes}
\MakeOuterQuote{"}

\begin{document}

\title{Quantum and Semi-Quantum Lottery: Strategies and Advantages}

\author[1]{Sandeep Mishra}
\author[1,*]{ Anirban Pathak}
\affil[1]{Jaypee Institute of Information Technology, A 10, Sector 62, Noida,
UP-201309, India}
\affil[*]{Corresponding author: anirban.pathak@gmail.com}

\maketitle

\begin{abstract}
Lottery is a game in which multiple players take chances in the hope of getting some rewards in cash or kind. In addition, from the time of the early civilizations, lottery has also been considered as an apposite method to allocate scarce resources. Technically, any scheme for lottery needs to be fair and secure, but none of the classical schemes for lottery are unconditionally secure and fair. As fairness demands complete unpredictability of the outcome of the lottery, it essentially requires perfect randomness. Quantum mechanics not only guarantees the generation of perfect randomness, it can also provide unconditional security. Motivated by these facts, a set of strategies for performing lottery using different type of quantum resources (e.g., single photon states, and entangled states) are proposed here, and it's established that the proposed strategies leads to unconditionally secure and fair lottery schemes. A scheme for semi-quantum lottery that allows some classical users to participate in the lottery involving quantum resources is also proposed and the merits and demerits of all the proposed schemes are critically analysed. Its also established that the level of security is intrinsically related to the type of quantum resources being utilized. Further, its shown that the proposed schemes can be experimentally realized using currently available technology, and that may herald a new era of commercial lottery.
\end{abstract}

\section{Introduction} \label{Intro}

Lottery is a game of chances where multiple players hope for getting the rewards. The use of lotteries has been prevalent since the time of early human civilizations. For example, during the Roman empire, lotteries were used as a form of amusement for giving gifts to the guests \cite{blanche1950lotteries}. The first commercial application of the lottery was done by Roman Emperor Augustus Caesar as an alternative to increase in taxes for the purpose of funding the infrastructure projects of the city \cite{rubner1966economics}. In the modern history, the first officially recorded state controlled lottery was organized by Queen Elizabeth I in 1556-59 for the purpose of funding a set of projects \cite{ball2018gambling}. In this particular case, $4,00,00$ tickets of $0.5$ pound each were issued with a reward of $5,000$ pounds to the winner. Since then, the lotteries have globally evolved as a mechanism that states can adopt to collect money (without levying higher taxes) required for various people-centric projects.  Thus, the lottery is historically used for socially meaningful purposes, but it has close resemblance with the gambling and it can always be viewed as a kind of gambling. Consequently, nowadays commercial lottery with financial rewards is not considered righteous in many countries.

Despite the above issue, we are interested in the lottery as the use of lottery is not restricted to gambling and collection of funds by the states. In fact, lotteries have many other applications in the diverse fields \cite{boyce1994allocation}. In the modern theory of allocation of resources, there are primarily four ways of allocation namely merit, queue, auction and lotteries. In "merit", the persons are allotted  points based on many parameters and the entity with maximum points get the first preference. In "queue", the entity which has submitted the application first gets the reward. In "auction'', one who is willing to pay the maximum gets the rights. Of all the above, lotteries does a randomization and the entity gets hold of the resource purely by luck. In fact, lottery is the only method which is free from any type of inherent bias. Debates are going on  to find an optimum method for the allocation of resources \cite{shaddy2021use}, but  the lottery is often considered as a better and fairer way of allocating the scarce resources among the large number of applicants \cite{stone2007lotteries,saunders2008equality}. Specially, if the number of indivisible goods ($k$) is lesser than the number of applicants ($n$) then lottery is considered as a suitable way for the allocation.  Tracing down in history, the lottery has been used by governments for the allotment of lands to the farmers. Even nowadays lotteries are used for the allocation of low cost houses by many governments across the world. Lotteries are prevalent medium in many countries for fair grant of admission to the students in the elementary schools. Lotteries are also used to grant work permits from the pool of eligible applicants. Lotteries are considered as a good way of placing the teams in various groups of major sporting events such as Olympics and the events organized by FIFA, NBA, etc. Even there is a recorded history of use lottery in the legal system where punishment to the accused were delivered via use of lottery in a situation where the act of crime was committed by a mob and it was difficult to trace out the right person who dealt the fatal blow \cite{eckhoff1989lotteries}. With respect to technology, lottery based CPU scheduling among the various competing processes has been proposed and used for instantaneous fair CPU allocation \cite{waldspurger1995lottery}.  Indeed, recently, lottery ticket hypothesis for graph theory has been proposed for training of the neural networks \cite{frankle2018lottery}. Nowadays, serious debate is also going on for the use of lotteries for the funding of the research projects as the prevalent medium of the peer review process has many intrinsic biases \cite{liu2020acceptability}. In fact, many funding agencies such as the Health Research Council of New Zealand, Volkswagen Foundation in Germany and the Swiss National Science Foundation are using the lotteries to fund the research projects after the initial screening of the eligible projects \cite{fang2016grant, adam2019science}.

As described above, lottery is an integral part of many important processes that are associated with our daily life. However, not all forms of lottery can be considered as fair. To understand this point, we first need to understand the meaning of a fair lottery. A lottery is considered as fair if and only if all the participants have an equal chance of winning. Thus, it requires perfect randomness. Further, once the results are announced then no one should be able to forge the ticket and claim to be the winner. Moreover, every participant should be able to verify the outcome of the process. An important point of concern is that the fairness of the lottery is inherently dependent upon the security of the lottery scheme being used. Thus, randomness and security are the primary concerns associated with the schemes of the lottery. Currently, the lottery schemes being used depend upon the credibility of the trusted authorities or security based on some mathematical complexities. Security derived in such a way is conditional. In fact, an unconditional security cannot be obtained in the classical word. Further, randomness used in classical schemes is weak compared to the randomness that can be generated quantum mechanically. Naturally, often issues have been raised regarding the fairness of the lottery schemes  and such things will come up again and again until and unless we have an unconditionally secure lottery scheme. By unconditionally secure, we mean that any potential adversary even with the unlimited resources would not be able to manipulate the outcome. Such issues were raised for the classical cryptographic schemes, too, but the advent of quantum cryptography  \cite{bennett1984quantum} provided a new way forward for unconditionally secure cryptography \cite{shenoy2017quantum,gisin2002quantum,xu2020secure}. The use of quantum states is currently being explored for providing unconditional security in various applications such as bit commitment \cite{brassard1993qbc,lo1997qbc,mayers1997qbc}, auctions \cite{qa1,qa2}, voting \cite{hillery2006,vaccaro2007,thapliyal2017qv,mishra2021qv}, multi-party computation \cite{colbeck2009}. Lottery is inherently related to quantum states as quantum mechanics has intrinsic randomness and lottery demands a complete randomization of the outcome \cite{danilov2018dynamic}. 

Quantum strategies for fair and unconditionally secure lottery is a demand of the time. A step in this direction was provided by Sun et al. in Ref. \cite{sun2020lottery}, where they proposed the schemes for lottery and auction on the backbone of the quantum blockchain. However, the mentioned protocol was not mature as it was based on quantum bit commitment which  still does not provide unconditional security. Further, it  used  the elementary idea of quantum blockchain where communication between nodes was done via the QKD protocol for which trust between the nodes would have been required at the forefront. However, the blockchain requires consensus on the contents of the decentralized data between between non-trusting parties. Barring this work, the field for use of quantum systems in implementing lottery schemes has remained largely untouched till now. Hence, we try to explore the use of quantum resources towards the development of fair and unconditionally secure lottery schemes which can be implemented via the currently available quantum hardware.

The rest of the paper is structured as follows. In Section \ref{def}, we introduce some of the basic ideas and nomenclature required for better understanding of the article with specific attention to the requirements that a good scheme of lottery should satisfy. Subsequently, a set of schemes for quantum and semi-quantum lottery are proposed in Section \ref{schemes}. This is followed by security  analysis of the proposed schemes in  Section \ref{security}. Finally, the paper is concluded in Section \ref{conclusion}.

\section{Basic notations and definitions }\label{def}

The basic requirements to be satisfied by a lottery scheme can be briefly mentioned as follows:

\begin{enumerate}[label=\roman*)]

\item \textbf{Eligibility:} Only the registered and legitimate entities can take part in the lottery.

\item \textbf{Equi-probability:} All the entities have equal probability to win the lottery. Thus, if there are $n$ participants, then the probability of winning for every participant ($p_i$) must be the same (complete randomization) and the total probability should be equal to unity. i.e.,
\begin{equation}
p_1=p_2=p_3= \dots = p_n = \mathcal{P}, \; \; \; \textrm{s.t.} \; \sum_{i=1}^{n} p_i =1.
\end{equation}

\item \textbf{Binding:} No one can change the lottery ticket after it has been issued.

\item \textbf{Verifiability:} Anyone can verify the outcome of the process.

\item \textbf{Secure:} An adversary even with unconditional computational power cannot manipulate the outcome.

\end{enumerate}

\section{Quantum lottery schemes} \label{schemes}

\begin{figure}[tb]
    	\includegraphics[width=\textwidth]{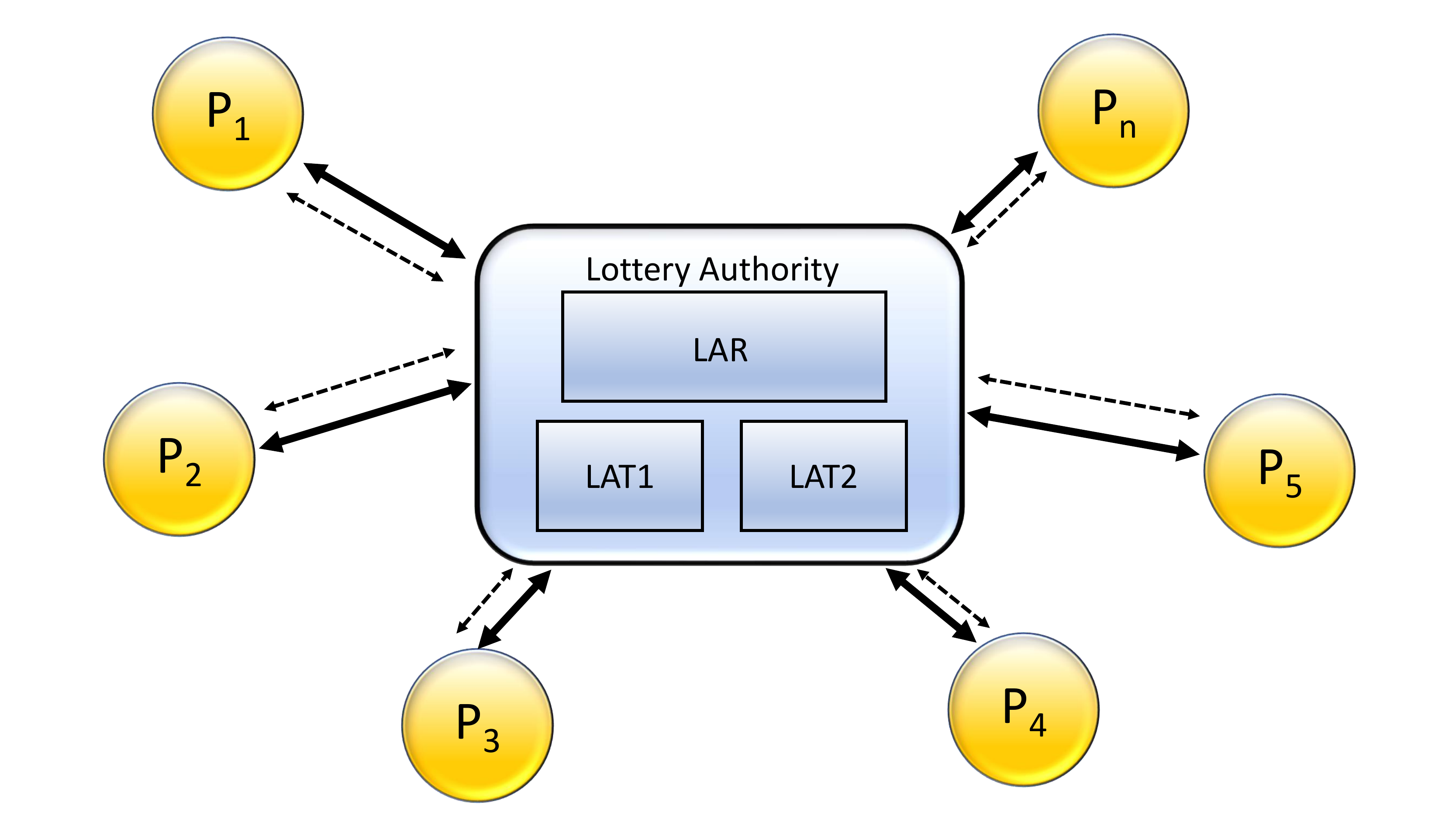}
    	\caption{(Color online) Schematic of the quantum lottery scheme with solid lines denoting quantum channels while dashed lines denoting classical channels. } 
    	\label{fig1}
	\end{figure}

The proposed lottery schemes consist of the following stakeholders (see fig \ref{fig1}):

1. \textbf{Lottery Authority:} The lottery authority ($LA$) is responsible for the conduct of lottery. Further, it will consist of multiple personnels but for the sake of simplicity, we can consider it to be consisting of three agents only, namely `lottery authority for registration' ($LAR$), `lottery authority for ticketing 1' ($LAT1$) and `lottery authority for ticketing 2' ($LAT2$). The role of $LAR$ is to register the interested parties and record their details. The role of $LAT1$ and $LAT2$ is to generate the lottery tickets for every eligible participant. Further, they cooperate with each other to declare the winning lottery ticket. 

2. \textbf{Participants:} Participants ($P_i$) consist of the set of people who are interested to participate in the lottery. It is to be mentioned that the proposed lottery scheme  provides a method  by which every participant can verify the winning lottery ticket.

\subsection{BB84 based quantum lottery scheme}

It has always been an endeavour since ages to develop a scheme by which secret messages can be sent from one party to another with minimum number of assumptions. The classical schemes were basically based on the assumption of trust that the encryption key is available to only authorized people or that some problems are too complex to be solved in polynomial time \cite{menezes2018handbook}. However, the advent of quantum cryptography in 1984 (i.e., the introduction of BB84 protocol for quantum key distribution (QKD)) \cite{bennett1984quantum} altogether changed the rules of the game by providing a scheme for unconditionally secure distribution of keys. Further, Ekert showed that the unconditional secure distribution of keys can be done via the use of entanglement \cite{ekert1991quantum} and  the presence of unauthorized interceptor can be detected by checking the correlations between entangled particles. This lead to two different ways of secure key distribution and each has its own advantages and disadvantages. Currently, this field of unconditional secure cryptography is quite mature to be used in practical situations and scenarios  \cite{shenoy2017quantum,gisin2002quantum,pan2020qkdreview,pirandola2020advances}. Parallel to the development of quantum key distribution technology, researchers have been working on the development of true random number generators whose outputs are completely non-deterministic as well as private. Since quantum mechanics has intrinsic randomness associated with it \cite{bera2017randomness}, so the quantum systems can be used to generate truly random numbers \cite{ananthaswamy2019turn}. Currently, the quantum technology is quite mature that we can generate very high quality random numbers at great speeds \cite{mannalath2022comprehensive,jacak2021quantum,herrero2017quantum,ma2016quantum} for a wide variety of  applications including secure communication, e-commerce, multi-party computations and lottery. In fact, there are many commercial quantum random number generator (QRNG) devices available in the market such as ID Quantique \cite{QuantisQ20:online}, Toshiba \cite{httpswww1:online}, PicoQuant \cite{QuantumR34:online}, MPD \cite{MicroPho90:online} etc.  

Taking inspiration from quantum cryptographic protocols, and the availability of the required hardware, we will first propose a lottery strategy based on BB84 states \cite{nielsen} and then briefly elaborate about its physical implementation. BB84 states are one of the most important and widely studied set of states studied in context of quantum key distribution. These states can be experimentally produced in a wide variety of physical systems, but it is more useful when implemented using photonic systems. In photonics, BB84 states are essentially polarization-encoded qubits or equivalently a sequence of photons which are randomly polarized in horizontal, vertical, $45^{\rm{o}}$ or $-45^{\rm{o}}$ which respectively correspond to the states $\vert0\rangle$, $\vert 1 \rangle$, $\vert + \rangle$ and $\vert - \rangle$ \cite{pathak_book}. These states can be easily generated by passing the photons from a laser diode source to an attenuator (neutral density filter) or by performing heralding on the output of certain spontaneous parametric down conversion process, and then passing the single photon through the relevant polarizer \cite{pathak2016optical}. The lottery scheme using BB84 states involves the following three stages:

\subsubsection{Registration phase \label{bb84_registration}} 

This phase is required for the registration of every participant with the $LA$ and generation of unique digital signatures similar to that proposed by Wallden et al. \cite{wallden2015quantum} for every participant $P_i$. The steps involved in the process can be enumerated as follows: 
\begin{description}

\item[{Step~1}] The participant $P_i$ will first register with the $LA$ by sending the documents and purchasing the lottery ticket with $LAR$. 

\item[{Step~2}] $LAR$ will verify the credentials of the $P_i$ and then use a QRNG to  issue a unique  256 bit participant's identity ($PID$) for the participant $P_i$. Further, $LAR$ keeps a record of all the allocated $PID$s in the database. The purpose of generation of $PID$ for every participant is to maintain the privacy of the participants as from here on the participant will only be using $PIDs$ in all the subsequent steps. Because of the use of QRNG by $LAR$, the probability of $PID$ collision for two participants will be asymptomatically very small. However, if the $PID$ generated for a new participant collides with the existing set of allocated $PID$s, then the QRNG is used again to generate a new $PID$.   

\item[{Step~3}] For the generation of digital signatures, the participant $P_i$  will generate two large identical, but random sequences of BB84 states ($\vert 0 \rangle,\vert 1 \rangle,\vert + \rangle,\vert - \rangle$) in accordance with the output generated by QRNG. $P_i$ will then send the first and second sequence respectively to $LAT1$ and $LAT2$ via the use of quantum channel. 

\item[{Step~4}] $LAT1$ ($LAT2$) will randomly choose to either forward the signature element to  $LAT2$ ($LAT1$) or keep it with themselves to directly measure it. Further, in either case the position of the elements in the sequence is recorded.

\item[{Step~5}] $LAT1$ ($LAT2$) measures the states that they have directly received from $P_i$ or through $LAT2$ ($LAT1$) by randomly choosing either the rectilinear basis ($\vert 0 \rangle,\vert 1 \rangle$) or diagonal basis ($\vert + \rangle,\vert - \rangle$). In this way, after the measurements, both $LAT1$ and  $LAT2$ exclude at least one of the four possible states and generate an eliminated signature for the sequence. e.g., if the measurement result is  $\vert 0 \rangle$ then  the participant must have never sent $\vert 1 \rangle$. The eliminated signature will serve as the quantum digital signature for the participant $P_i$ to be used in the next phase. 

\end{description}

Similarly, unconditionally secure digital signatures are generated for every participant $P_i$.  The registration phase in only meant for the generation of signatures for every participant that will be used in the next phase for the authentication of the participants before the generation of the tickets.

\subsubsection{Ticketing phase: \label{bb84_ticketing}}  

In this phase, lottery ticket numbers are generated by every participant. The steps involved in this phase are as follows: 
\begin{description}

\item[{Step~1}] Participant $P_i$ will first send the $PID$ to both $LAT1$ and $LAT2$. After that, $P_i$ will reveal the classical information corresponding to the BB84 sequence used in the registration stage.

\item[{Step~2}] Both $LAT1$ and $LAT2$ will then match the set of states revealed by $P_i$ with the corresponding eliminated signatures for every position. If the number of mismatches as recorded by either of $LAT1$ and $LAT2$ is greater than a threshold limit, then the participant is not allowed to take part further. $P_i$ is allowed to participate only if he is authenticated by both  $LAT1$ and $LAT2$. 

\item[{Step~3}] $P_i$ will use any of the experimentally available BB84 based QKD protocol  to generate two keys, namely $K_{P_i}^{LAT1}$ and $K_{P_i}^{LAT2}$ corresponding to $LAT1$ and $LAT2$.

\item[{Step~4}] $P_i$ will then generate a random 256-bit unique ticket ($TID_i$) via use of QRNG. These $TID$s will be used for the draw of lots. Also, the hash value of the $TID_i$ will be generated and publicly announced. The generation of $TID_i$ by the participant will prevent any kind of manipulation by the lottery authority during ticket allocation phase. Further, the public announcement of hash of $TID_i$ precludes any forging of the lottery ticket after the reward have been announced. Here, it is to be mentioned that any adversary can use the dictionary attack in which the publicly announced hash value of $TID_i$ can be compared with the pre-calculated hash value of all possible 256 bit $TID$s. But, such an attack is computationally impossible. 

\item[{Step~5}] $P_i$ sends the $TID_i$ to both $LAT1$ and $LAT2$ using the key  $K_{P_i} = K_{P_i}^{LAT1} \oplus K_{P_i}^{LAT2}$. In this way, the $TID$ sent by $P_i$ can be opened only if $LAT1$ and $LAT2$ cooperate with each other. As an alternative, $P_i$ can use any other experimentally available quantum secret sharing protocol using BB84 states \cite{guo2003quantum,hsu2005quantum} to send the $TID_i$ to $LAT1$ and $LAT2$.

\end{description}

The same procedure will be repeated by every participant $P_i$ to send the correspondingly generated $TID_i$ to the $LAT1$ and $LAT2$. No lottery tickets will be accepted after the closing of the phase.

\subsubsection{ Rewards phase: \label{bb84_rewards}}  The steps involved in this phase are as follows:

\begin{description}

\item[{Step~1}] $LAT1$ and $LAT2$ cooperate with each other to open the tickets $TID_i$.

\item[{Step~2}] The winning ticket is announced as the bit wise XOR of all the received $TIDs$. i.e., $T^W= \oplus TID_i$.

\item[{Step~3}] The reward for every participant is calculated as proportional to the Hamming distance of the participants’ ticket ($TIDi$) from the winning ticket (TW). Hamming distance between two strings basically corresponds to the number of positions in which two strings are different. For bit strings, it corresponds to the number of 1’s present in the XOR of the two strings. In the proposed scheme, there is finite possibility that two or more participants have the same $TID$. For such an event, the rewards can be distributed equally to all participants having the winning ticket, but for most of the cases the specific scheme for the calculation of rewards will depend upon the particular application of the lottery scheme. 

\end{description}

In this phase, our main focus is to just propose a method for generation of the winning ticket rather than commenting on a particular method for distribution of the rewards. The allocation of rewards will depend upon the application of lottery scheme and is thus left open to the potential users to develop their own methods. Further, such methods only constitute as a part of the post-processing techniques.

	\begin{figure}[tb]
    	\includegraphics[width=\textwidth]{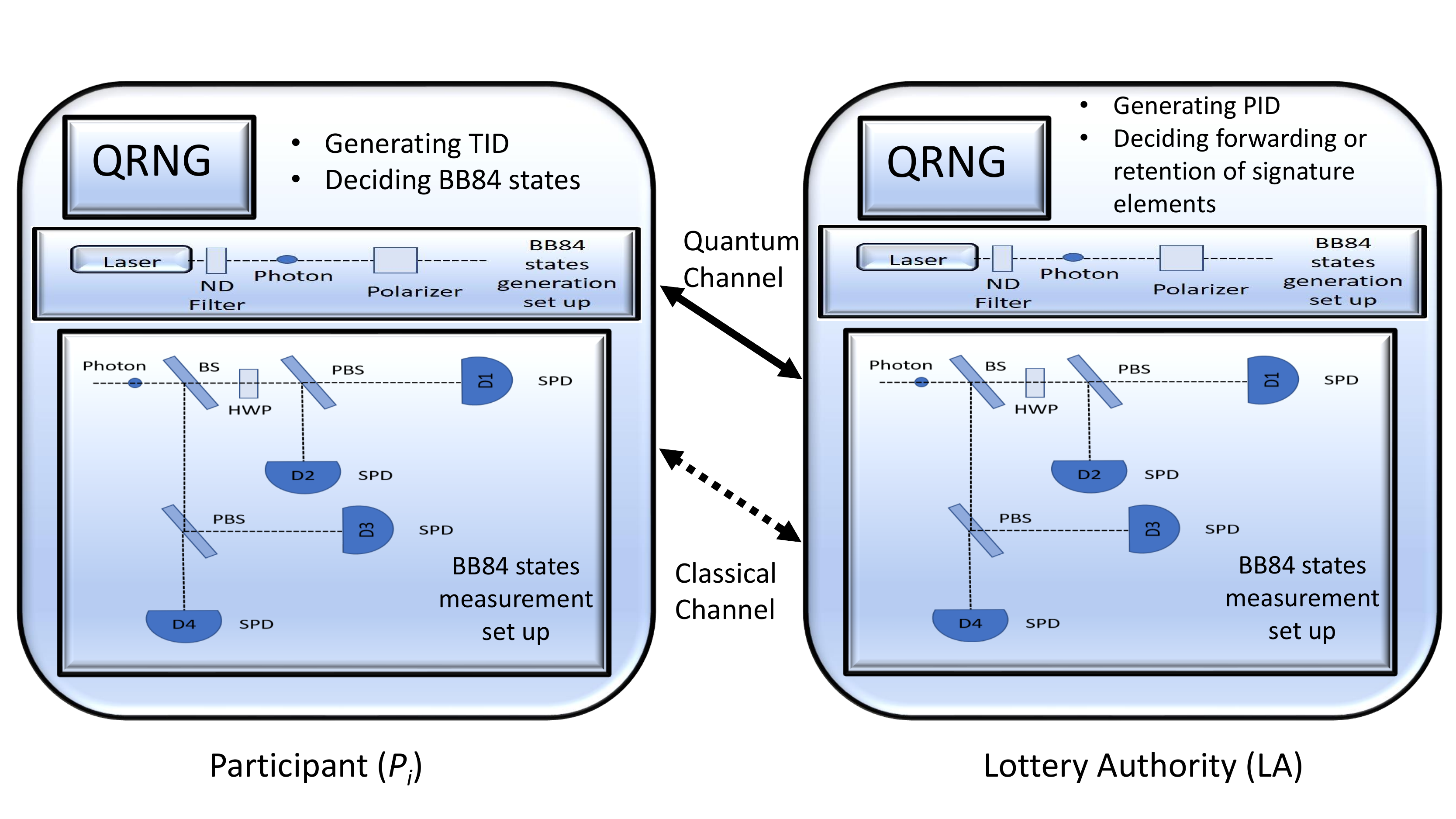}
    	\caption{(Color online) Schematic of the resource requirements for BB84 based lottery scheme. All the participants as well as lottery authority require a QRNG, a set up for generation of BB84 states and a set up for measurement of BB84 states.} 
    	\label{fig2}
	\end{figure}

Before, moving to the next lottery scheme, let us just briefly describe the physical implementation of the above mentioned scheme. Figure \ref{fig2} shows the schematic of the resource requirements for the participants and the lottery authority to implement the BB84 based lottery scheme. We can see that the resource requirements at both the ends are symmetric in nature. Both of them require a QRNG to perform certain steps in the protocol and more so with respect to generation of TIDs and PIDs. As mentioned before, currently there are a wide variety of commercial QRNGs in the market, though the cost is a bit on the higher end. To reduce cost, the participants may use pseudo random number generators (PRNGs) which pass NIST test \cite{nist} but that may lead to a compromise with the security aspects. Other than QRNG, both of them require set-up for the generation and the measurement of BB84 states. The current optical technology is quite mature enough to generate and measure BB84 states with very high fidelity. For photonics based implementation, the qubits are encoded in the polarization states of the single photon with  horizontal, vertical, $45^{\rm{o}}$ or $-45^{\rm{o}}$ respectively corresponding to the states $\vert0\rangle$, $\vert 1 \rangle$, $\vert + \rangle$ and $\vert - \rangle$ \cite{pathak_book}. These BB84 states are used for both the authentication of the participants as well as sending of the TIDs from the participants to the lottery authority. Further, the participants and lottery authority are connected to each other via bi-directional quantum as well as the classical authenticated channel. Current technology now allows us to send qubits from one party to another through free space communication as well as via use of optical fibres. So, the proposed lottery scheme seems feasible to be implemented via the use of currently available technology.

\subsection{Entanglement based quantum lottery scheme}

Till now, we have proposed a lottery scheme based on BB84 states, but this is not the only quantum system that can used. In fact, we already know that for quantum cryptography, there are many entanglement based protocols \cite{ekert1991quantum,pirandola2020advances,pan2020qkdreview}. The use of entanglement brings altogether new features such as device independence \cite{lo2012measurement,yin2016measurement,colbeck2009}. i.e. we need not trust the devices used for the implementation. Similar to the use of entanglement in cryptography, here we will propose a lottery scheme by exploiting the feature of entanglement that can be found only in quantum systems. It is to be mentioned that quantum entanglement is a very costly resource which is very difficult to maintain. So, we will try to minimize the use of quantum entanglement in the proposed scheme. We will keep the registration phase and rewards phase same as that used in the already discussed scheme while using the entanglement only in the ticketing phase using schemes similar to that of quantum secret sharing \cite{hillery1999quantum,zhang2005multiparty,xiao2004efficient}. This is done with the motivation that ticketing is the most important phase where tickets are generated by the participants and are sent to the lottery authority.  The proposed scheme makes use of the Bell state $\vert \psi^{-} \rangle= \frac{1}{\sqrt{2}} \{ \vert 01 \rangle - \vert 10 \rangle\}$ and single qubit local unitary operators, namely $U_{0}= \vert0\rangle\langle0\vert + \vert1\rangle\langle1\vert $, $U_{1}= \vert0\rangle\langle0\vert - \vert1\rangle\langle1\vert $, $U_{2}= \vert1\rangle\langle0\vert + \vert0\rangle\langle1\vert $, $U_{3}= \vert0\rangle\langle1\vert - \vert1\rangle\langle0\vert $. In principle, the scheme can use any of the four Bell states ($\vert \psi^{\pm} \rangle, \vert \phi^{\pm} \rangle$) \footnote{$\vert \psi^{\pm} \rangle= \frac{1}{\sqrt{2}} \{ \vert 01 \rangle \pm \vert 10 \rangle\}$ and $\vert \phi^{\pm} \rangle= \frac{1}{\sqrt{2}} \{ \vert 00 \rangle \pm \vert 11 \rangle\}$} but we will use only $\vert \psi^{-} \rangle$ in the scheme. Further, the scheme will make use of the entanglement swapping for two Bell states \cite{zukowski1993event,ji2022entanglement}. The steps involved in the ticketing phase while implementing entanglement based lottery scheme are as follows:

\begin{description}
\item[{Step~1}] same as that of \ref{bb84_ticketing}
\item[{Step~2}] same as that of \ref{bb84_ticketing}

\item[{Step~3}] $P_i$ will prepare 256 pair of Bell states $\vert \psi^{-} \rangle$ and stores the first qubit of all 256 pairs with him. The sequence of second qubits of one set of 256 Bell states is sent to $LAT1$ while the sequence of second qubits of the other set of 256 Bell states is sent to $LAT2$. So, $P_i$ shares 256 Bell states with $LAT1$ (Set I) and another 256 Bell states with $LAT2$ (Set II). The combined state of $P_i$, $LAT1$ and $LAT2$ can be written as $\{ \vert \psi^{-} \rangle^{1}\otimes \vert \psi^{-} \rangle^{2}\otimes \dots \otimes\vert \psi^{-} \rangle^{256} \}^{P_i}_{LAT1} \bigotimes \{ \vert \psi^{-} \rangle^{1}\otimes \vert \psi^{-} \rangle^{2}\otimes \dots \otimes\vert \psi^{-} \rangle^{256} \}^{P_i}_{LAT2} $.

\item[{Step~4}] $P_i$ randomly picks half of his qubits, then choose to measure them either in the computational basis ($\vert 0 \rangle, \vert 1 \rangle$) or Haddamard basis ($\vert + \rangle, \vert - \rangle$) and publicly announce the  basis used for measurement of his qubits. $LAT1$ and $LAT2$ will use the same basis as announced by $P_i$ to measure their corresponding qubits. If there is no adversary during the transmission phase, then the measurement outcomes of $P_i$ will be opposite to that of $LAT1$. e.g. If $P_i$ gets $\vert 0\rangle$ ($\vert 1\rangle$) then $LAT1$ will get $\vert 1\rangle$ ($\vert 0\rangle$) while if $P_i$ gets $\vert + \rangle$ ($\vert - \rangle$) then $LAT1$ will get $\vert - \rangle$ ($\vert +\rangle$).  Similar is the case for measurement outcomes of $P_i$ and $LAT2$. If the error is below the threshold limit, then they proceed to next step else they abort the protocol. Further, they will rearrange the qubits to have the combined state as $\{ \vert \psi^{-} \rangle^{1}\otimes \vert \psi^{-} \rangle^{2}\otimes \dots \otimes\vert \psi^{-} \rangle^{128} \}^{P_i}_{LAT1} \bigotimes \{ \vert \psi^{-} \rangle^{1}\otimes \vert \psi^{-} \rangle^{2}\otimes \dots \otimes\vert \psi^{-} \rangle^{128} \}^{P_i}_{LAT2} $.

\item[{Step~5}] $P_i$ will choose one qubit from Set I and another qubit from Set II. Further, $P_i$ will randomly choose one of the above qubits to apply any one of the operators $U_0, U_1, U_2, U_3$ to encode the bits $00, 01, 10, 11$ respectively. $P_i$ will then perform a Bell measurement on the two qubits and then publicly announce the outcome. The same operation is performed for all the qubits of $P_i$ by taking one qubit from Set I while other from Set II. In this way, $P_i$ will send the encoded 256 bit $TID$ to $LAT1$ and $LAT2$. Further, the hash of the $TID$ is publicly announced.

\item[{Step~6}] $LAT1$ and $LAT2$ can cooperate with each other to get the $TID$ sent by $P_i$ after performing the necessary joint Bell measurements of their respective qubits and the properties of entanglement swapping.
 
\end{description}

In this way, all the $P_i$s will send their $TID$s to the $LA$ and the $TID$s can be opened only if both the $LAT1$ and $LAT2$ cooperate with each other. So, entanglement based lottery scheme differs from that of BB84 based scheme only in the method adopted to send the ticketing numbers ($TID$s) from participants to the lottery authority. The advantage of using entanglement is just to provide an additional level of security.

\begin{figure}[tb]
    	\includegraphics[width=\textwidth]{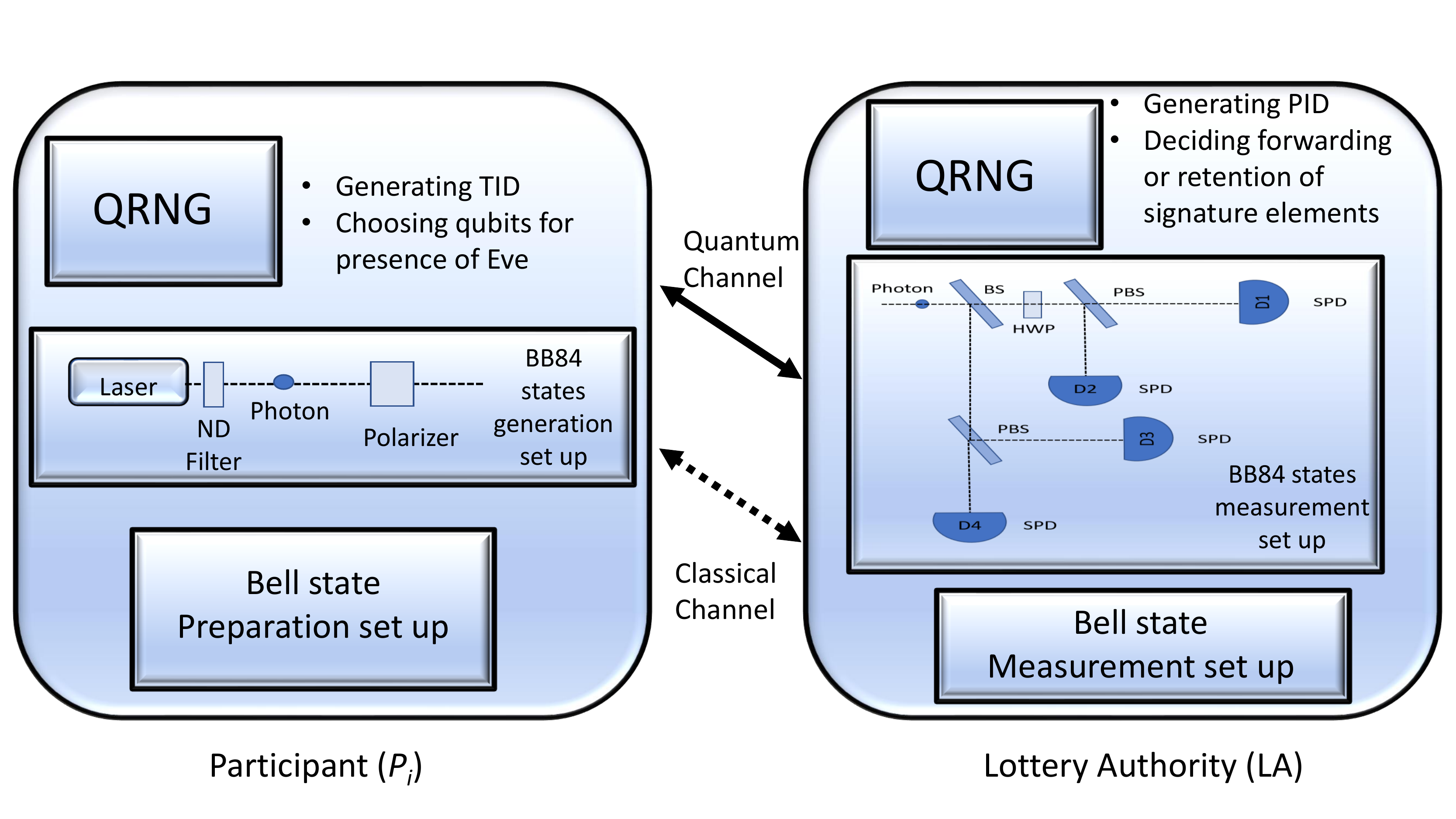}
    	\caption{(Color online) Schematic of entanglement based lottery scheme. Participants require a BB84 state generation set up while $LA$ require a BB84 state measurement set up for registration phase while ticketing phase requires Bell state preparation set up with participants and Bell state measurement set up with $LA$. Other than this, participants and $LA$ also require QRNG.} 
    	\label{fig3}
	\end{figure}

Let us now briefly describe the physical implementation of the proposed scheme. Figure \ref{fig3} shows the schematic of the resource requirements for  participants and the lottery authority to implement the entanglement  based lottery scheme. In contrast to the BB84 based scheme, the resource requirements here are not symmetric with respect to the participants and the lottery authority except the use of QRNGs at both the ends. As mentioned before, in this scheme too authentication of the participants is done via the use of BB84 based digital signature scheme. But for its implementation, only the BB84 state generation set-up is required at the participants end while the BB84 state measurement set-up is required by the lottery authority. Further, the TIDs are sent from participants to the lottery authority via encoding the TIDs into the Bell states. So, the Bell state generation set up is required by the participants while Bell state measurement set up is required  by lottery authority. Similar to the case of BB84 based scheme, participants and the lottery authority are connected to each other via a quantum channel and classical authenticated channel. As far as current technology is concerned, the proposed scheme can be physically implemented, but certainly the use of fragile resources such as entanglement will come at a very heavy cost. In the next section, we will propose a scheme to minimize the cost by allowing all participants to use classical   resources, while only lottery authority having access to quantum resources.

\subsection{Semi-quantum lottery scheme}

In the previous two proposed schemes, all the participants need to have the quantum capabilities. However, in the realistic situations the quantum resources are extremely costly and difficult to maintain. In fact, entanglement is a very fragile resource  too. In order to overcome these limitations, Boyer et al. \cite{boyer2007} in 2007 proposed a semi-quantum scheme of quantum key distribution,  in which only one party has full quantum abilities but the other party is classical. The classical party can either reflect back the qubits or can measure the incoming qubits in the computational basis ($\vert0\rangle,\vert1\rangle$) only. Since then, various semi-quantum protocols have been proposed in quantum cryptography \cite{zou2009semiquantum,yu2014authenticated,shukla2017semi} and related areas. As far a current scenario is concerned, infrastructure for  classical communication systems is very well developed and is available at a very reasonable cost. But in contrast, the quantum resources such as creation and  manipulation of quantum states, quantum entanglement are too costly and difficult to handle. So, the current situation demands the development of schemes in which only few nodes have full quantum capabilities while the rest of the nodes can make of use only classical resources. Such schemes are known as semi-quantum schemes, and those schemes are relevant as they can exploit advantages of the currently available classical infrastructure. Taking the above motivation, we will now present a semi-quantum lottery scheme in which lottery authority will have full quantum capability, but all the participants will have only classical abilities. The proposed semi-quantum lottery scheme is further shown to be equally good in comparison with their quantum counterparts. Let us now describe in detail the semi-quantum scheme for lottery.

\subsubsection{Registration phase:} 

\begin{description}

\item[{Step~1}] same as that of \ref{bb84_registration}  

\item[{Step~2}] same as that of \ref{bb84_registration}

\item[{Step~3}] For the generation of digital signatures, $LAR$ will prepare $n$ qubits in the state $\vert + \rangle$ and send it to $P_i$ one by one. The participant $P_i$ can either measure the incoming qubit in the computational basis ($\vert0\rangle,\vert 1 \rangle$) or let it go as it is to $LAT1$.

\item[{Step~4}] $LAT1$ will randomly choose to measure the qubit in either computational basis ($\vert0\rangle,\vert 1 \rangle$) or Hadamard basis ($\vert + \rangle,\vert - \rangle$) and note the outcome. After performing the measurement,  resultant qubit is sent to the participant $P_i$.

\item[{Step~5}] The participant $P_i$ will perform exactly the same operation as done by him in Step 3 (i.e., either pass the qubit or measure in computational basis) on incoming qubit from $LAT1$ and send it to $LAT2$. 

\item[{Step~6}] Similar to $LAT1$, $LAT2$ will also randomly choose either computational basis ($\vert0\rangle,\vert 1 \rangle$) or Hadamard basis ($\vert + \rangle,\vert - \rangle$)  to measure the qubit and note the outcome.  

\item[{Step~7}] $LAT1$ and  $LAT2$ will announce the basis used by them to measure each of the $n$ qubits received by them. Further, they will keep only those outcomes in which they have used the same basis and discard the rest. These outcomes will be used by $LAT1$ and $LAT2$ to verify the participant in the ticketing phase. 

\end{description}

\subsubsection{Ticketing phase:} 

\begin{description}

\item[{Step~1}] Participant $P_i$ will first send the $PID$ to $LAT1$ and $LAT2$. After that, $ P_i$ will reveal the information on all $n$ qubits, whether he has allowed the qubit to pass to $LAT1$ and $LAT2$ or measured in the computational basis before sending them to $LAT1$ and $LAT2$.

\item[{Step~2}] Both $LAT1$ and $LAT2$ will then match the outcomes of their measurement  for the cases in which participant $P_i$ has not measured the qubit before passing it to them. For such cases, the outcome recorded by $LAT1$ and $LAT2$ will match with each other. If the number of mismatches  is greater than a threshold limit, then the participant is not allowed to take part further. $P_i$ is allowed to participate only if he is authenticated by  $LAT1$ and $LAT2$.

\item[{Step~3}] $P_i$ will use the semi-quantum QKD scheme using BB84 protocol given by Boyer et al. \cite{boyer2007} or any other semi-quantum QKD protocol to generate two keys namely $K_{P_i}^{LAT1}$ and $K_{P_i}^{LAT2}$ corresponding to $LAT1$ and $LAT2$.

\item[{Step~4}] same as that of \ref{bb84_ticketing}

\item[{Step~5}] $P_i$ sends the $TID_i$ to both $LAT1$ and $LAT2$ using the key  $K_{P_i} = K_{P_i}^{LAT1} \oplus K_{P_i}^{LAT2}$ or via use of any other semi-quantum quantum secret sharing protocol. In this way, the $TID$ sent by $P_i$ can be opened only if $LAT1$ and $LAT2$ cooperate with each other.

\end{description}

\subsubsection{Rewards phase:} same as that of \ref{bb84_rewards}

\begin{figure}[tb]
    	\includegraphics[width=\textwidth]{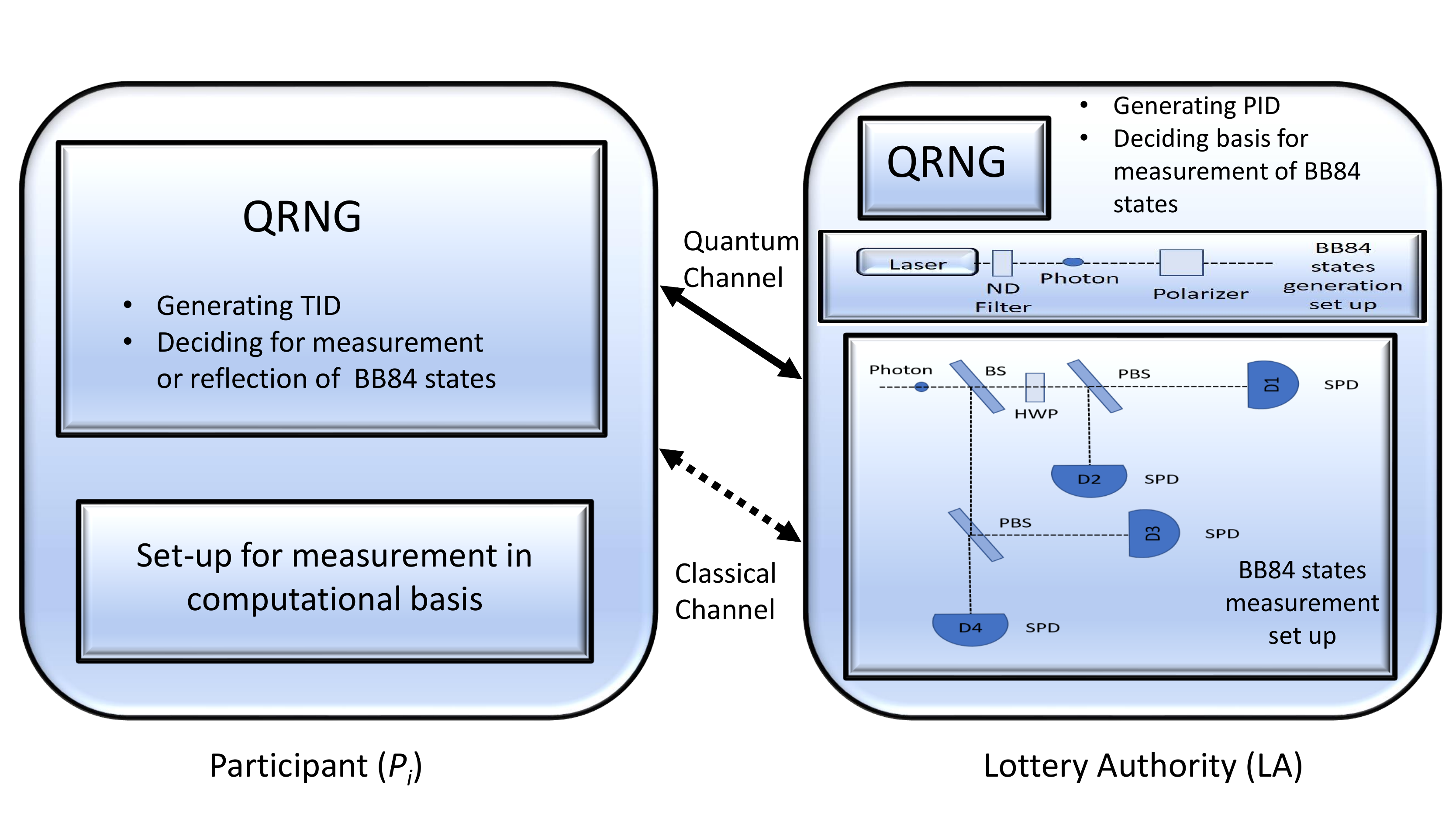}
    	\caption{(Color online) Schematic of semi-quantum based lottery scheme. Participants require a QRNG and a set up for measurement of BB84 states in computational basis. $LA$ is fully quantum and requires a QRNG, a set up for BB84 state generation and a set up for measurement of BB84 states.} 
    	\label{fig4}
	\end{figure}

So, we can see that the lottery scheme can be implemented using the lesser resources in comparison to that of full quantum lottery schemes. It is to be mentioned here that the above lottery scheme is also a BB84 state based scheme, but this allows the users to have only classical resources with only lottery authority having full quantum capabilities. Figure \ref{fig4} describes the schematic of the resource requirements for participants and lottery authority to implement the protocol. We can clearly see that,  except for the QRNG (required for performing certain steps), participants just need to have access to quantum channel coming out from the lottery authority. After receiving the qubits from the lottery authority, the participants just need a set up for measurement of the incoming qubits in the computational basis ($\vert 0 \rangle, \vert 1 \rangle$). Also, they can allow the qubits to be returned back to lottery authority as it is without any modification. Further, to reduce the costs the participants may use PRNGs which passes the NIST tests, but at the cost of security. Looking at the lottery authority, they require quantum resources such as QRNG, BB84 states generation as well as measurement set-up. With the current advancements of technology, the proposed protocol can be implemented with the currently available hardware.

So far we have proposed three protocols for lottery, which can be realized using different amount (and type) of quantum resources. Further, we have shown that the proposed theoretical schemes can be implemented using the currently available hardware. However, the security of the protocols is not discussed in detail until now. Consequently, it will be apt to perform security analysis of the proposed lottery schemes in the next section.

\section{Security analysis} \label{security}

The proposed schemes conform to the requirements of a good lottery scheme as described in section \ref{def}. Also, in all the proposed schemes, we can see that the tickets are generated by the participants and the lottery authority only collects the generated $TID$s which are then further used for deciding the rewards. In this section, we will first look at the security aspects of the BB84 or entanglement based schemes (henceforth referred to as type I scheme) and then look at the security of semi-quantum schemes (henceforth referred to as type II scheme). For the type I scheme, we can see that after the registration every participant generates a digital signature with lottery authority by the use of unconditionally secure BB84 based digital signature scheme. These digital signatures are used for the authentication of the eligible participants and only eligible participants are allowed to generate the tickets and take part further in the process. So, the eligibility condition is satisfied. Now, since the winning ticket is announced by taking the bit wise XOR of all the valid $TID$s, so every $TID$ has an equal probability to win. Further, no one will be able to predict the winning ticket beforehand until and unless one gets hold of all the TIDs. So, the equi-probability condition is satisfied. 

The $TID$s are sent by the participants to the lottery authority using the unconditionally secure experimentally feasible BB84 protocol. So, it is practically impossible for any adversary to manipulate or change the ticket ID. Further, the $TID$s can be opened only if $LAT1$ and $LAT2$ cooperate with each other. Even in the entanglement based scheme, the $TID$s are sent to the lottery authority via the use of Quantum Secret Sharing (QSS) protocol. In this way, the tickets are sent from participants to lottery authority in an unconditionally secure way. Further, every participant publicly announces the hash value of their $TID$ before being sent to the lottery authority. This is done to prevent the participant to claim a different $TID$ once the rewards have been announced.  In this, the type I lottery schemes are unconditionally secure while sending of the ticket Id but the binding property of the $TID$ via hash function makes it only computationally secure as an adversary can perform the dictionary attack. In this attack, an adversary can compute the hash of all the $TID$s before hand and then via the mapping of the $TID$ with their hash value the adversary can get hold of all the transmitted $TID$. But since one is using 256 bit $TID$ so it is computationally not feasible for the current set of computers to do a mapping of all possible $TID$s with their hash value.  Now the type I scheme is verifiable as after the announcement of the winning ticket, every participant can announce their $TID$ which can be used to verify the outcome of the lottery. Further, any malicious participant can not change his ticket after the announcement of the winning ticket as every participant has announced the hash value of their $TID$ before being sent to the lottery authority. So, to conclude, the proposed type I lottery schemes satisfy all the requirements to be considered as good schemes. 

Let us now look at the security aspects of semi-quantum lottery scheme. The semi-quantum lottery scheme (type II) differs from the type I schemes in terms of resources as in the type II scheme only lottery authority needs quantum resources while the participants can be classical only. Due to this, the registration phase in which digital signatures for participants are generated is different from that of type I scheme. As can be seen from the scheme, the participants signature is encoded in the form of whether the participant has measured the incoming qubit in the computational basis or let it be passed without any modification. Lottery authority can authenticate the participant when he reveals his choices and $LAT1$ and $LAT2$ announce the measurement results. For all the qubits in which participant has passed the qubits, the measurement outcomes of $LAT1$ and $LAT2$ will be same provided they use the same basis. In this, only authenticated participants will be allowed to proceed in the ticketing phase. Further, in the ticketing phase the $TID$s are sent by participants to lottery authority via the use of semi-quantum QKD protocols which have already been proved to be unconditionally secure. So, the semi-quantum lottery scheme also satisfies the requirements of a good lottery scheme.

\section{Conclusion} \label{conclusion}

The Nature is quantum mechanical and the quantum mechanical world is probabilistic in nature. In short, quantum mechanics is a probabilistic theory and in our daily life we often come across situations that can be best realized within the framework of a probabilistic theory (not essentially quantum mechanics). Lottery is one such phenomenon which can be appropriately realized only in the framework of a probabilistic theory. It's possible to design schemes of lottery in any non-classical probabilistic theory. However, without going into the details of the generalized probability theory (GPT)\cite{janotta2014generalized,barrett2007information} and the specific toy theories which can support secure lottery schemes, here we have restricted ourselves to quantum mechanics and have provided three specific schemes for lottery. These schemes require different type of quantum resources. To be precise, in contrast to the entanglement based scheme proposed above, the other two schemes, i.e., BB84 state based scheme and semi-quantum schemes can be realized using separable states or more appropriately using single photon states. In fact, these two single-photon based (or equivalently BB84 state based) schemes do not require entanglement and non-locality, and thus such schemes can also be realized in a non-classical toy theory  \cite{aravinda2019hierarchical} which has only the feature of uncertainty relations between incompatible observables. This is similar to the availability of a wide variety of practical QKD systems \cite{pan2020qkdreview,pirandola2020advances} and quantum random number generators \cite{mannalath2022comprehensive,herrero2017quantum} with different levels of security aspects. The unconditional security can be derived from the use of only `incompatibility and uncertainty' feature of quantum mechanics, but the device independence security can be derived only through the use of features such as `entanglement' and `non-locality'.
 
In this work, we have highlighted the importance of lottery in many important works of life and have noted that despite its existence from the days of early civilization there is no fair and secure scheme for lottery. In fact, current implementations of the schemes for lottery are not fully secure. Further, we identified the requirements for a scheme to be considered as a good scheme for lottery. The gap is addressed here by designing a set of secure schemes for lottery and establishing their security. We analysed the resource requirements of the three proposed scheme, and have shown that these schemes, can be implemented using the available devices. We hope that this study will help in providing physical insights towards the development of commercial quantum lottery solutions.

\small

\section*{Acknowledgements} Authors thank Kishore Thapliyal and Abhishek Parakh for their feedback and interest in this work. Further, the authors acknowledge that this study is supported by the project ``\textit{Partnership 2020: Leveraging US-India Cooperation in Higher Education to Harness Economic Opportunities and Innovation}" which is enabling a collaboration between University of Nebraska at Omaha, and JIIT, Noida.

\footnotesize

\bibliographystyle{unsrt}

\bibliography{ql}

\begin{thebibliography}{10}

\bibitem{blanche1950lotteries}
Ernest~E Blanche.
\newblock Lotteries yesterday, today, and tomorrow.
\newblock {\em The Annals of the American Academy of Political and Social
  Science}, 269(1):71--76, 1950.

\bibitem{rubner1966economics}
Alex Rubner.
\newblock {\em The economics of gambling}.
\newblock London, Macmillan, 1966.

\bibitem{ball2018gambling}
Patrick~Seymour Ball.
\newblock {\em Gambling in Elizabethan England: perspectives on England’s
  ‘Lotterie Generall’of 1567--69}.
\newblock PhD thesis, University of Tasmania, 2018.

\bibitem{boyce1994allocation}
John~R Boyce.
\newblock Allocation of goods by lottery.
\newblock {\em Economic Inquiry}, 32(3):457--476, 1994.

\bibitem{shaddy2021use}
Franklin Shaddy and Anuj~K Shah.
\newblock When to use markets, lines, and lotteries: How beliefs about
  preferences shape beliefs about allocation.
\newblock {\em Journal of Marketing}, 2021.

\bibitem{stone2007lotteries}
Peter Stone.
\newblock Why lotteries are just.
\newblock {\em Journal of Political Philosophy}, 15(3):276--295, 2007.

\bibitem{saunders2008equality}
Ben Saunders.
\newblock The equality of lotteries.
\newblock {\em Philosophy}, 83(3):359--372, 2008.

\bibitem{eckhoff1989lotteries}
Torstein Eckhoff.
\newblock Lotteries in allocative situations.
\newblock {\em Social Science Information}, 28(1):5--22, 1989.

\bibitem{waldspurger1995lottery}
Carl~A Waldspurger.
\newblock {\em Lottery and stride scheduling: Flexible proportional-share
  resource management}.
\newblock PhD thesis, Massachusetts Institute of Technology, 1995.

\bibitem{frankle2018lottery}
Jonathan Frankle and Michael Carbin.
\newblock The lottery ticket hypothesis: Finding sparse, trainable neural
  networks.
\newblock In {\em International Conference on Learning Representations}, 2019.

\bibitem{liu2020acceptability}
Mengyao Liu, Vernon Choy, Philip Clarke, Adrian Barnett, Tony Blakely, and Lucy
  Pomeroy.
\newblock The acceptability of using a lottery to allocate research funding: a
  survey of applicants.
\newblock {\em Research Integrity and Peer Review}, 5:3, 2020.

\bibitem{fang2016grant}
Ferric~C Fang and Arturo Casadevall.
\newblock Grant funding: Playing the odds.
\newblock {\em Science}, 352(6282):158--158, 2016.

\bibitem{adam2019science}
David Adam.
\newblock Science funders gamble on grant lotteries.
\newblock {\em Nature}, 575(7785):574--575, 2019.

\bibitem{bennett1984quantum}
Charles~H Bennett and G~Brassard.
\newblock Quantum cryptography: {P}ublic key distribution and coin tossing.
\newblock In {\em International Conference on Computer System and Signal
  Processing, IEEE, 1984}, pages 175--179, 1984.

\bibitem{shenoy2017quantum}
Akshata Shenoy-Hejamadi, Anirban Pathak, and Srikanth Radhakrishna.
\newblock Quantum cryptography: {K}ey distribution and beyond.
\newblock {\em Quanta}, 6(1):1--47, 2017.

\bibitem{gisin2002quantum}
Nicolas Gisin, Gr{\'e}goire Ribordy, Wolfgang Tittel, and Hugo Zbinden.
\newblock Quantum cryptography.
\newblock {\em Reviews of Modern Physics}, 74(1):145--195, 2002.

\bibitem{xu2020secure}
Feihu Xu, Xiongfeng Ma, Qiang Zhang, Hoi-Kwong Lo, and Jian-Wei Pan.
\newblock Secure quantum key distribution with realistic devices.
\newblock {\em Reviews of Modern Physics}, 92(2):025002, 2020.

\bibitem{brassard1993qbc}
Gilles Brassard, Claude Cr{\'e}peau, Richard Jozsa, and Denis Langlois.
\newblock A quantum bit commitment scheme provably unbreakable by both parties.
\newblock In {\em Proceedings of 1993 IEEE 34th Annual Foundations of Computer
  Science}, pages 362--371. IEEE, 1993.

\bibitem{lo1997qbc}
Hoi-Kwong Lo and Hoi~Fung Chau.
\newblock Is quantum bit commitment really possible?
\newblock {\em Physical Review Letters}, 78(17):3410--3413, 1997.

\bibitem{mayers1997qbc}
Dominic Mayers.
\newblock Unconditionally secure quantum bit commitment is impossible.
\newblock {\em Physical Review Letters}, 78(17):3414--3417, 1997.

\bibitem{qa1}
Mosayeb Naseri.
\newblock Secure quantum sealed-bid auction.
\newblock {\em Optics communications}, 282(9):1939--1943, 2009.

\bibitem{qa2}
Tad Hogg, Pavithra Harsha, and Kay-Yut Chen.
\newblock Quantum auctions.
\newblock {\em International Journal of Quantum Information}, 5(05):751--780,
  2007.

\bibitem{hillery2006}
Mark Hillery, M{\'a}rio Ziman, Vladim{\'\i}r Bu{\v{z}}ek, and Martina
  Bielikov{\'a}.
\newblock Towards quantum-based privacy and voting.
\newblock {\em Physics Letters A}, 349(1-4):75--81, 2006.

\bibitem{vaccaro2007}
Joan~Alfina Vaccaro, Joseph Spring, and Anthony Chefles.
\newblock Quantum protocols for anonymous voting and surveying.
\newblock {\em Physical Review A}, 75(1):012333, 2007.

\bibitem{thapliyal2017qv}
Kishore Thapliyal, Rishi~Dutt Sharma, and Anirban Pathak.
\newblock Protocols for quantum binary voting.
\newblock {\em International Journal of Quantum Information}, 15(01):1750007,
  2017.

\bibitem{mishra2021qv}
Sandeep Mishra, Kishore Thapliyal, Abhishek Parakh, and Anirban Pathak.
\newblock Quantum anonymous veto: A set of new protocols.
\newblock {\em arXiv preprint arXiv:2109.06260}, 2021.

\bibitem{colbeck2009}
Roger Colbeck.
\newblock Quantum and relativistic protocols for secure multi-party
  computation.
\newblock {\em PhD thesis, University of Cambridge, arXiv:0911.3814}, 2009.

\bibitem{danilov2018dynamic}
Vladimir~Ivanovitch Danilov, Ariane Lambert-Mogiliansky, and Vassili
  Vergopoulos.
\newblock Dynamic consistency of expected utility under non-classical (quantum)
  uncertainty.
\newblock {\em Theory and Decision}, 84(4):645--670, 2018.

\bibitem{sun2020lottery}
Xin Sun, Piotr Kulicki, and Mirek Sopek.
\newblock Lottery and auction on quantum blockchain.
\newblock {\em Entropy}, 22(12):1377, 2020.

\bibitem{menezes2018handbook}
Alfred~J Menezes, Paul~C Van~Oorschot, and Scott~A Vanstone.
\newblock {\em Handbook of applied cryptography}.
\newblock CRC press, 2018.

\bibitem{ekert1991quantum}
Artur~K Ekert.
\newblock Quantum cryptography based on {B}ell’s theorem.
\newblock {\em Physical Review Letters}, 67(6):661--663, 1991.

\bibitem{pan2020qkdreview}
Feihu Xu, Xiongfeng Ma, Qiang Zhang, Hoi-Kwong Lo, and Jian-Wei Pan.
\newblock Secure quantum key distribution with realistic devices.
\newblock {\em Reviews of Modern Physics}, 92(2):025002, 2020.

\bibitem{pirandola2020advances}
Stefano Pirandola, Ulrik~L Andersen, Leonardo Banchi, Mario Berta, Darius
  Bunandar, Roger Colbeck, Dirk Englund, Tobias Gehring, Cosmo Lupo, Carlo
  Ottaviani, et~al.
\newblock Advances in quantum cryptography.
\newblock {\em Advances in Optics and Photonics}, 12(4):1012--1236, 2020.

\bibitem{bera2017randomness}
Manabendra~Nath Bera, Antonio Ac{\ifmmode\acute{\imath}\else\'{\i}\fi}n, Marek
  Ku{\ifmmode\acute{s}\else\'{s}\fi}, et~al.
\newblock {Randomness in quantum mechanics: philosophy, physics and
  technology}.
\newblock {\em Reports on Progress in Physics}, 80(12):124001, 2017.

\bibitem{ananthaswamy2019turn}
{How to Turn a Quantum Computer Into the Ultimate Randomness Generator
  {$\vert$} Quanta Magazine}, Jun 2019.
\newblock [Online; accessed 28. Feb. 2022].

\bibitem{mannalath2022comprehensive}
Vaisakh Mannalath, Sandeep Mishra, and Anirban Pathak.
\newblock A comprehensive review of quantum random number generators: Concepts,
  classification and the origin of randomness.
\newblock {\em arXiv preprint arXiv:2203.00261}, 2022.

\bibitem{jacak2021quantum}
Marcin~M. Jacak, Piotr J{\'o}{\'{z}}wiak, Jakub Niemczuk, and Janusz~E. Jacak.
\newblock Quantum generators of random numbers.
\newblock {\em Scientific Reports}, 11(1):16108, 2021.

\bibitem{herrero2017quantum}
Miguel Herrero-Collantes and Juan~Carlos Garcia-Escartin.
\newblock {Quantum random number generators}.
\newblock {\em Reviews of Modern Physics}, 89(1):015004, 2017.

\bibitem{ma2016quantum}
Xiongfeng Ma, Xiao Yuan, Zhu Cao, et~al.
\newblock {Quantum random number generation - npj Quantum Information}.
\newblock {\em npj Quantum Information}, 2(16021):1--9, 2016.

\bibitem{QuantisQ20:online}
ID~Quantique.
\newblock Quantis qrng (quantum random number generator) - id quantique.
\newblock \url{https://www.idquantique.com/random-number-generation/products}.
\newblock (Accessed on 25/02/2022).

\bibitem{httpswww1:online}
Toshiba.
\newblock
  https://www.toshiba.eu/pages/eu/cambridge-research-laboratory/quantum-random-number-generators.
\newblock
  \url{https://www.toshiba.eu/pages/eu/Cambridge-Research-Laboratory/quantum-random-number-generators}.
\newblock (Accessed on 25/02/2022).

\bibitem{QuantumR34:online}
PicoQuant.
\newblock Quantum random number generator | picoquant.
\newblock
  \url{https://www.picoquant.com/scientific/product-studies/pqrng-150-product-study}.
\newblock (Accessed on 25/02/2022).

\bibitem{MicroPho90:online}
MPD.
\newblock Micro photon devices - quantum random number.
\newblock \url{http://www.micro-photon-devices.com/Docs/Datasheet/QRNG.pdf}.
\newblock (Accessed on 25/02/2022).

\bibitem{nielsen}
Michael~A Nielsen and Isaac~L Chuang.
\newblock {\em Quantum {C}omputing and {Q}uantum {I}nformation}.
\newblock Cambridge University Press, Cambridge, 2000.

\bibitem{pathak_book}
Anirban Pathak.
\newblock {\em Elements of quantum computation and quantum communication}.
\newblock CRC Press Boca Raton, 2013.

\bibitem{pathak2016optical}
Anirban Pathak and Anindita Banerjee.
\newblock {\em Optical quantum information and quantum communication}.
\newblock SPIE, 2016.

\bibitem{wallden2015quantum}
Petros Wallden, Vedran Dunjko, Adrian Kent, and Erika Andersson.
\newblock Quantum digital signatures with quantum-key-distribution components.
\newblock {\em Physical Review A}, 91(4):042304, 2015.

\bibitem{guo2003quantum}
Guo-Ping Guo and Guang-Can Guo.
\newblock Quantum secret sharing without entanglement.
\newblock {\em Physics Letters A}, 310(4):247--251, 2003.

\bibitem{hsu2005quantum}
Li-Yi Hsu and Che-Ming Li.
\newblock Quantum secret sharing using product states.
\newblock {\em Physical Review A}, 71(2):022321, 2005.

\bibitem{nist}
Andrew Rukhin, Juan Soto, James Nechvatal, Miles Smid, and Elaine Barker.
\newblock A statistical test suite for random and pseudorandom number
  generators for cryptographic applications.
\newblock Technical report, Booz-allen and hamilton inc mclean va, 2001.

\bibitem{lo2012measurement}
Hoi-Kwong Lo, Marcos Curty, and Bing Qi.
\newblock Measurement-device-independent quantum key distribution.
\newblock {\em Physical Review Letters}, 108(13):130503, 2012.

\bibitem{yin2016measurement}
Hua-Lei Yin, Teng-Yun Chen, Zong-Wen Yu, Hui Liu, Li-Xing You, Yi-Heng Zhou,
  Si-Jing Chen, Yingqiu Mao, Ming-Qi Huang, Wei-Jun Zhang, et~al.
\newblock Measurement-device-independent quantum key distribution over a 404 km
  optical fiber.
\newblock {\em Physical Review Letters}, 117(19):190501, 2016.

\bibitem{hillery1999quantum}
Mark Hillery, Vladim{\'\i}r Bu{\v{z}}ek, and Andr{\'e} Berthiaume.
\newblock Quantum secret sharing.
\newblock {\em Physical Review A}, 59(3):1829, 1999.

\bibitem{zhang2005multiparty}
Zhan-jun Zhang, Yong Li, and Zhong-xiao Man.
\newblock Multiparty quantum secret sharing.
\newblock {\em Physical Review A}, 71(4):044301, 2005.

\bibitem{xiao2004efficient}
Li~Xiao, Gui~Lu Long, Fu-Guo Deng, and Jian-Wei Pan.
\newblock Efficient multiparty quantum-secret-sharing schemes.
\newblock {\em Physical Review A}, 69(5):052307, 2004.

\bibitem{zukowski1993event}
Marek Zukowski, Anton Zeilinger, Michael~A Horne, and Aarthur~K Ekert.
\newblock "{E}vent-ready-detectors" {B}ell experiment via entanglement
  swapping.
\newblock {\em Physical Review Letters}, 71(26):4287--4290, 1993.

\bibitem{ji2022entanglement}
Zhaoxu Ji, Peiru Fan, and Huanguo Zhang.
\newblock Entanglement swapping for bell states and
  {G}reenberger--{H}orne--{Z}eilinger states in qubit systems.
\newblock {\em Physica A: Statistical Mechanics and its Applications},
  585:126400, 2022.

\bibitem{boyer2007}
Michel Boyer, Dan Kenigsberg, and Tal Mor.
\newblock Quantum key distribution with classical {B}ob.
\newblock {\em Physical Review Letters}, 99(14):140501, 2007.

\bibitem{zou2009semiquantum}
Xiangfu Zou, Daowen Qiu, Lvzhou Li, Lihua Wu, and Lvjun Li.
\newblock Semiquantum-key distribution using less than four quantum states.
\newblock {\em Physical Review A}, 79(5):052312, 2009.

\bibitem{yu2014authenticated}
Kun-Fei Yu, Chun-Wei Yang, Ci-Hong Liao, and Tzonelih Hwang.
\newblock Authenticated semi-quantum key distribution protocol using bell
  states.
\newblock {\em Quantum Information Processing}, 13(6):1457--1465, 2014.

\bibitem{shukla2017semi}
Chitra Shukla, Kishore Thapliyal, and Anirban Pathak.
\newblock Semi-quantum communication: protocols for key agreement, controlled
  secure direct communication and dialogue.
\newblock {\em Quantum Information Processing}, 16(12):1--19, 2017.

\bibitem{janotta2014generalized}
Peter Janotta and Haye Hinrichsen.
\newblock {Generalized probability theories: what determines the structure of
  quantum theory?}
\newblock {\em Journal of Physics A: Mathematical and Theoretical},
  47(32):323001, 2014.

\bibitem{barrett2007information}
Jonathan Barrett.
\newblock {Information processing in generalized probabilistic theories}.
\newblock {\em Physical Review A}, 75(3):032304, 2007.

\bibitem{aravinda2019hierarchical}
S.~Aravinda, Anirban Pathak, and R.~Srikanth.
\newblock {Hierarchical axioms for quantum mechanics}.
\newblock {\em European Physical Journal D}, 73(9):207, 2019.

\end{thebibliography}

\end{document}